\documentclass{PoS}

\usepackage{dsfont}
\usepackage{psfrag}

\newcommand{\unitop}{\mathds{1}}

\newcommand{\Tr}[1]{\mathrm{Tr} \left[ \, #1 \, \right]}
\newcommand{\Real}[1]{\mathrm{Re}\Big( \, #1 \, \Big)}
\newcommand{\Imag}[1]{\mathrm{Im}\Big( \, #1 \, \Big)}
\newcommand{\bea}{\begin{eqnarray}}
\newcommand{\eea}{\end{eqnarray}}
\newcommand{\be}{\begin{equation}}
\newcommand{\ee}{\end{equation}}

\newcommand{\nn}{\nonumber}
\newcommand{\onehalf}{\frac{1}{2}}
\newcommand{\eqnref}[1]{eqn.~(\ref{#1})}

\newcommand{\figref}[1]{fig.~(\ref{#1})}

\newcommand{\tabref}[1]{tab.~(\ref{#1})}

\newcommand{\usersizeone}[1]{{#1}}

\def\BA{\begin{eqnarray}}
\def\BE{\begin{equation}}
\def\EA{\end{eqnarray}}
\def\EE{\end{equation}}

\def\gtsim{\lower-0.45ex\hbox{$>$}\kern-0.77em\lower0.55ex\hbox{$\sim$}}
\def\ltsim{\lower-0.45ex\hbox{$<$}\kern-0.77em\lower0.55ex\hbox{$\sim$}}

\title{Hamiltonian Lattice QCD near the Light Cone}

\ShortTitle{Hamiltonian Lattice QCD near the Light Cone}

\author{\speaker{H.J. Pirner}$^{ac}$
        , D. Gr\"unewald$^a$
        ,E.-M. Ilgenfritz$^b$
        and E.V. Prokhvatilov$^d$\\
        \llap{$^a$}Institut f\"ur Theoretische Physik, Universit\"at Heidelberg, Germany\\
        \llap{$^b$}Institut f\"ur Physik, Humboldt-Universit\"at zu Berlin, Germany\\
        \llap{$^c$}Max-Planck-Institut f\"ur Kernphysik Heidelberg, Germany\\
        \llap{$^d$}Department of Theoretical Physics, St. Petersburg University, Russia\\
        E-mail: \email{pirner@tphys.uni-heidelberg.de}, \\ 
                \email{d.gruenewald@tphys.uni-heidelberg.de}, \\
                \email{ilgenfri@physik.hu-berlin.de}, \\
                \email{Evgeni.Prokhvat@pobox.spbu.ru}}

\abstract{We give a status report of our work on light cone Hamiltonian lattice 
          QCD. We have derived an effective Hamiltonian $H_{eff}$ which is only 
          quadratic in the momenta and therefore can be simulated  by standard methods.
          For this Hamiltonian we determine variationally an approximate ground state 
          wave functional in the light cone limit.}

\FullConference{The XXV International Symposium on Lattice Field Theory\\
		 July 30-August 4,2007\\
		 Regensburg, Germany}

\begin{document}

\section{Introduction}

Although the Hamiltonian is not Lorentz invariant,
the light cone Hamiltonian \cite{pauli,Burkardt:2001jg} 
offers the advantage of being boost invariant and has
- naively interpreted -  
a trivial vacuum. On the other hand, one  would be surprised if QCD looses its 
nonperturbative vacuum structure in the light cone limit. Probably 
much of the complicated vacuum structure of QCD is hidden in
the constraint equations appearing in light cone QCD. 
Remarkable progress has been made in light cone QCD with a color dielectric 
lattice theory as a starting point \cite{Bardeen:1979xx, Dalley:2003uj}. This
approach is based on ``fat'' links which arise from averaging gluon
configurations \cite{Pirner:1991im}. 
With this method  the spectrum of glueballs and the pion light cone wave 
function have been calculated \cite{ Dalley:2003uj}.  On the light cone
one is prevented from approaching the continuum limit, since the effective 
Hamiltonian for the link matrices $M\in GL(N)$ approaching  $U\in SU(N)$ 
is not known. 
This is the reason why we propose to formulate QCD near the 
light cone using $SU(N)$ link variables. Near light cone time plays a similar role as ordinary Minkowski time,
therefore we can follow the conventional method of the transfer matrix.  
The transversal
field strengths are increased in magnitude due to the boost into the vicinity of the light 
cone whereas the longitudinal fields remain unchanged. Constraint equations arise in the light cone
Hamiltonian framework which enforce the ``equality'' of the transverse 
chromo-electric and chromo-magnetic fields $E_k^a=F_{-k}^a$.
The lattice Hamiltonian density depends on an effective constant which
represents the product $\tilde \eta = \xi\eta$  of the asymmetry parameter $\xi=a_-/a_{\bot}$
and the near light cone parameter $\eta$. If one chooses $\eta=1$ and lets $\xi \rightarrow 0$ 
one obtains a deformed system which is squeezed in the spatial $a_-$ direction, if one uses $\xi=1$
and lets $\eta \rightarrow 0$ one obtains the light cone limit. This 
equivalence has been found  by Verlinde and Verlinde
\cite{Verlinde:1993te} and Arefeva \cite{Arefeva:1993hi}.
These authors have proposed to implement 
such asymmetric lattices in order to study high energy scattering. This has
motivated us to proceed further in this way.  
In the work of Balitsky \cite{Balitsky:2001gj} one approaches the 
light cone from time-like distances which is close to the scattering experiments. 
However, we approach the light cone from space-like distances.
The asymmetric lattice Hamiltonian itself is not usable for Monte Carlo methods 
since the electric
field strengths i.e. the momenta appear linearly. Because of the
translational invariance of the vacuum we can add a term $1/\eta^2 P_-$ 
to cancel the unwanted terms. Naively this amounts to returning to an
effective lattice 
Hamiltonian which is proportional to the energy in ordinary Minkowski coordinates. 
This is a reasonable procedure to search the groundstate of the vacuum. Applications of the
light cone coordinates in finite temperature field theory have followed the same
route \cite{Raufeisen:2004dg}.

\section{The QCD Hamiltonian near the light cone}

We introduce near light
cone (nlc) coordinates, first proposed by
\cite{Prokhvatilov:1989eq}  
\BA
x^+&=&\frac{1}{\sqrt2}\left[\left(1+\frac{\eta^2}{2}\right)x^{
0}+ \left(1-\frac{\eta^2}{2}\right)x^{ 3}\right] \nonumber\\
x^-&=&\frac{1}{\sqrt2}\left[x^{0}-x^{3}\right] \; .
\label{eqn:TrafoNLC}
\EA
The transversal coordinates $x^1$ and $x^2$ remain unchanged. 
The near light cone parameter $\eta$ may be interpreted as parameterizing a Lorentz boost 
into a frame which is moving with velocity $\beta=(1-\eta^2/2)/(1+\eta^2/2)$ along the longitudinal 
direction relative to the laboratory frame. 
Then, the nlc energy $p_+$ and longitudinal momentum $p_-$ expressed in terms of the 
laboratory energy $E_{lab}$ and longitudinal momentum $p^3_{lab}$ are given by
\bea
p_+&=& \frac{1}{\eta}\left(E_{lab}-p^3_{lab}\right)\nn\\ 
p_-&=& \eta \, p^3_{lab}~.                      
\label{eqn:MomentaRel}
\eea 
The second relation in \eqnref{eqn:MomentaRel} shows that 
large longitudinal momenta $p^3_{lab}$ become
accessible by a nlc lattice with a cut-off $p_-\propto 1/a_-$.

The definition of 
nlc coordinates \eqnref{eqn:TrafoNLC} induces the following metric:
\begin{equation}
 \begin{array}{cc}
  g_{\mu\nu}=\left(
   \begin{array}{cccc}
      0 &  0 &  0 &  1 \\
      0 & -1 &  0 &  0 \\
      0 &  0 & -1 &  0 \\
      1 &  0 &  0 &  -\eta^2 
   \end{array}
             \right)  &
  g^{\mu\nu}=\left(
   \begin{array}{cccc}
     \eta^2 &  0 &  0 &  1 \\
          0 & -1 &  0 &  0 \\
          0 &  0 & -1 &  0 \\
          1 &  0 &  0 &  0
   \end{array}
             \right) 
   \end{array}
\label{eq.(2)}   
\end{equation}
with $\mu,\nu=+,1,2,-,\det g=1$. This defines the scalar product
\BA
x_\mu y^\mu & = & x^-y^+ + x^+y^- - \eta^2 x^-y^- - \vec x_\perp\vec
y_\perp\nonumber\\ 
            & = & x_-y_+ + x_+y_- + \eta^2 x_+y_+ - \vec x_\perp\vec
y_\perp \; .
\label{eq.(3)}
\EA
Note, that the metric has off-diagonal terms which implies that there are
terms mixing temporal and longitudinal spatial coordinates in the scalar product.
This yields a form of the pure gluonic Lagrange density \eqnref{eqn:LagrangeDensnlc} which has
severe consequences for a numerical treatment
\bea \mathcal{L}&=&
\sum\limits_a \left[ \onehalf F_{+-}^aF_{+-}^a
  +\sum\limits_{k=1}^2\left(F_{+k}^aF_{-k}^a+
    \frac{\eta^2}{2}F_{+k}^aF_{+k}^a \right) -\onehalf
  F_{12}^aF_{12}^a \right] \; .
\label{eqn:LagrangeDensnlc}                     
\eea 
The Lagrange density is 
linear in one of the temporal field strengths, namely
$F_{+k}^aF_{-k}^a$. Therefore, a standard Monte Carlo sampling of the Euclidean path
integral does not work for nlc coordinates and we 
rather use a Hamiltonian formulation.

The energy momentum tensor in its most general form is given by
\bea
T^{\mu\nu}&=&\sum\limits_r 
\frac{\delta \mathcal{L}}{\delta\left(\partial_{\mu} \Phi_r\right)} 
\partial^{\nu} \Phi_r - g^{\mu\nu} \mathcal{L} \; .
\eea
It defines the Hamiltonian density $\mathcal{H}=T_{\;\;\;+}^+ \;$ and the longitudinal momentum density 
$\mathcal{P}_-=T_{\;\;\;-}^+$ 
as
\bea 
\mathcal{H} &=& \onehalf \sum\limits_a 
\left[ \Pi_-^{a}\Pi_-^{a}+F_{12}^aF_{12}^a
+ \sum\limits_{k=1}^2 \frac{1}{\eta^2}\left(\Pi_k^{a}-F_{-k}^a\right)^2
                                       \right] \nn\\
\mathcal{P}_-&=& \Pi_-^a \partial_- A_-^a +
\sum\limits_{k=1}^2 \Pi_k^a \partial_- A_k^a  \; .
\eea 
This form of the local integrand for the generator $\mathcal{P}_-$ of longitudinal translations 
is not manifestly gauge invariant. 
However, if one uses Gauss' law and the definition of the field 
\-strengths tensors one can rewrite $\mathcal{P}_-$ in a symmetrized form
\bea
\mathcal{P}_-&=& \onehalf\left(\Pi_k^a F_{-k}^a+F_{-k}^a\Pi_k^a\right) \; .
\eea
In order to solve the Hamiltonian we are
interested in translation-invariant states which are eigenstates of
the longitudinal momentum operator, i.e.  with eigenvalue equal zero.
In vacuum, with light cone momentum $P_-=0$, we can add
$(1/\eta^2)~P_-$ to define an effective Hamiltonian density
$\mathcal{H}_{eff}$ which is only quadratic in momenta: 
\bea
\mathcal{H}_{eff}&=& \mathcal{H}+\frac{1}{\eta^2}\mathcal{P}_- \nn \\
&=& \frac{1}{2} \sum\limits_a \left[
  \Pi_-^{a}\Pi_-^{a}+F_{12}^aF_{12}^a + \sum\limits_{k=1}^2
  \frac{1}{\eta^2}\left(\Pi_k^{a}\Pi_k^{a} + F_{-k}^aF_{-k}^a\right)
\right] \; .
\label{eq:quadratic}
\eea 
In a forthcoming paper we show how to derive \cite {pirner} the effective lattice
Hamiltonian with the coupling constant $\lambda=4/g^4$ using the transfer matrix:
\bea
\mathcal{H}_{\mathrm{eff,lat}}&=& \frac{1}{N_-N_{\bot}^2}\frac{1}{a_{\bot}^4} \frac{2}{\sqrt{\lambda}}
\sum\limits_{\vec{x}}\left\{
    \frac{1}{2}~\sum\limits_{a}~{\Pi}_-^a(\vec{x})^2~
    +\onehalf~\lambda~\Tr{\unitop-\Real{{U}_{12}(\vec{x})}}
    \right.  \nn\\
& &  \left.
    ~+\sum\limits_{k,a}~\frac{1}{2}~\frac{1}{\tilde{\eta}^2}~ 
      \Bigg[~{\Pi}_k^a(\vec{x})^2+
      \lambda~\Biggl(\Tr{\frac{\sigma_a}{2}~\Imag{U_{-k}(\vec{x})}}\Biggr)^2
      \Bigg] 
    \right\}\; .
\label{eqn:EffectiveLatHamiltonian}    
\eea
For $\tilde{\eta}=1$ this effective
Hamiltonian is very similar to the traditional Hamiltonian used in
equal time lattice theory. It differs only in the potential energy terms
for the $U_{-k}$ plaquettes. Instead of the usual
$\mathrm{Tr}[\unitop-\mathrm{Re}(U_{-k})]$ term resembling the field strength squared
in the naive continuum limit, the nlc Hamiltonian has the form
 $(\mathrm{Tr}[\sigma^a/2\;\mathrm{Im}(U_{-k})])^2$ which 
corresponds to the plaquette in the adjoint representation and which yields an  
additional $Z(2)$ symmetry in comparison with the full lattice Hamiltonian \cite{pirner}. 
The light cone limit $\tilde{\eta} \rightarrow 0$ 
enhances the importance of transverse
electric and magnetic fields without 
generating unwanted linear terms in the momenta. The resulting vacuum solution
should be a plausible extrapolation of
the vacuum solution of  QCD.
 

\section{Variational optimization of the ground state wavefunctional}

We analytically solve the effective lattice Hamiltonian for the ground state
wavefunctional both in the strong and weak coupling limit \cite{pirner}. Both solutions
can be described by a product of single plaquette wavefunctionals for $\tilde{\eta}$ 
sufficiently close to one. In order to cover
the whole coupling range, we make a variational Ansatz of the ground state 
wavefunctional which is given by a product of single 
plaquette wavefunctionals with two variational parameters $\rho$ and $\delta$ and 
with the normalization constant N
\bea
\Psi_0(\rho,\delta)&=& N \prod\limits_{\vec{x}}\exp\left\{\sum\limits_{k=1}^2 
                 \rho\,\Tr{\Real{U_{-k}(\vec{x})}}
                +\delta\,\Tr{\Real{U_{12}(\vec{x})}}
               \right\}~. \nn\\
           & & \phantom{1}    
\label{eqn:VariationlGSWF}
\eea
With this normalized wavefunction we optimize the 
energy expectation value $\epsilon_0(\rho,\delta)$ of the effective 
Hamiltonian for fixed values of the coupling $\lambda$ and the near
light cone parameter $\tilde{\eta}$.
\begin{figure}[ht]
	\centering
	    \psfrag{x1\r}{\usersizeone{$\lambda$}}
	    \psfrag{x2\r}{\usersizeone{$\rho_0/\lambda^{1/2}$}}
	    \psfrag{t1111111111\r}{\usersizeone{$\tilde{\eta}^2=1$}}
	    \psfrag{t2222222222\r}{\usersizeone{$\tilde{\eta}^2=0.2$}}
	    \psfrag{t3333333333\r}{\usersizeone{$\tilde{\eta}^2=0.05$}}
	    \includegraphics[width=1.0\textwidth]{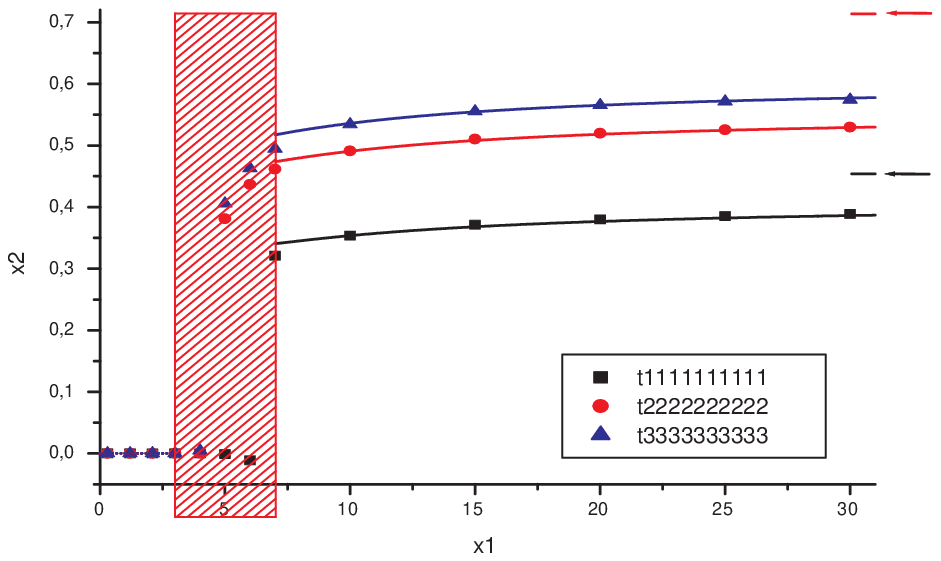}
	\caption{Optimal wavefunctional parameter $\rho_0(\lambda,\tilde{\eta})$ as a function of $\lambda$ 
	         obtained from the simulation on a $N_\perp^2\times N_-=16^3$ lattice for three different values 
	         of $\tilde{\eta}^2$.
	         The red shaded area corresponds to the phase transition region for all values of $\tilde{\eta}^2$. 
	         The dotted lines show the predicted analytical strong coupling behavior. The arrows indicate
	         the expected asymptotic behavior for weak coupling which is proportional to $\sqrt{\lambda}$, i.e. a 
	         constant independent of $\lambda$ in the plot. 
	         The solid lines show the actual analytic parameterizations in the weak coupling regime (c.f. 
	         eqn. (3.3)).
	         \label{fig:Params_first_16x16x16}} 
\end{figure}
\begin{figure}[ht]
	\centering
	    \psfrag{x1\r}{\usersizeone{$\lambda$}}
	    \psfrag{x2\r}{\usersizeone{$\delta_0/\lambda^{1/2}$}}
	    \psfrag{t1111111111\r}{\usersizeone{$\tilde{\eta}^2=1$}}
	    \psfrag{t2222222222\r}{\usersizeone{$\tilde{\eta}^2=0.2$}}
	    \psfrag{t3333333333\r}{\usersizeone{$\tilde{\eta}^2=0.05$}}
	    \includegraphics[width=1.0\textwidth]{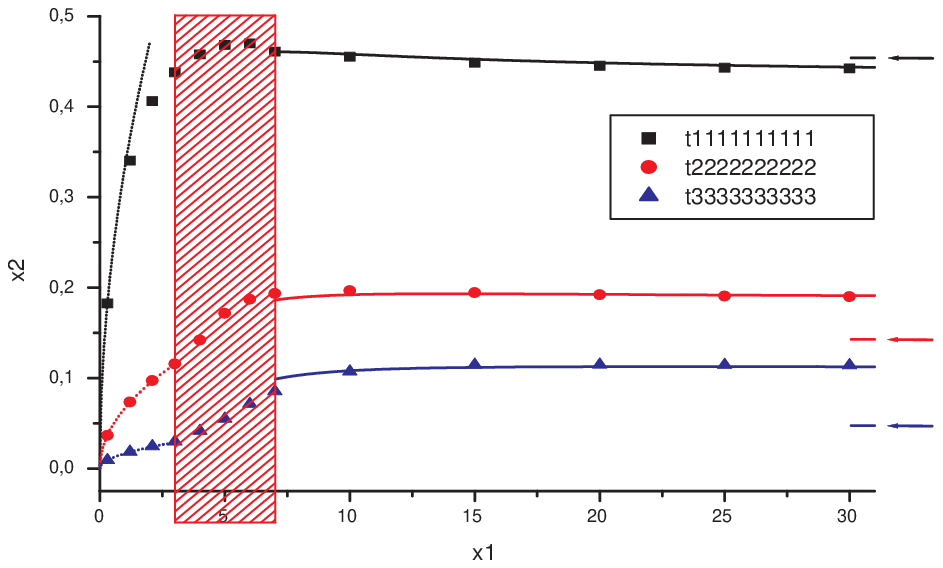}
	\caption{Optimal wavefunctional parameter $\delta_0(\lambda,\tilde{\eta})$ as a function of $\lambda$ 
	         obtained from the simulation on a $N_\perp^2\times N_-=16^3$ lattice for three different values of 
	         $\tilde{\eta}^2$.
	         The red shaded area corresponds to the phase transition region for all values of $\tilde{\eta}^2$. 
	         The dotted lines show the predicted analytical strong coupling behavior. The arrows indicate
	         the expected asymptotic behavior for weak coupling which is proportional to $\sqrt{\lambda}$, i.e. a 
	         constant independent of $\lambda$ in the plot. 
	         The solid lines show the actual analytic parameterizations in the weak coupling regime (c.f. 
	         eqn. (3.3)).
	         \label{fig:Params_second_16x16x16}} 
\end{figure}

In \figref{fig:Params_first_16x16x16} and \figref{fig:Params_second_16x16x16}, 
we present the variationally optimized wavefunctional
parameters $\rho_0$ and $\delta_0$ rescaled by a factor $1/\sqrt{\lambda}$ such
that they approach a constant in the assymptotic weak coupling region, i.e. $\lambda \rightarrow \infty$.
The statistical uncertainty on the variational parameters is typically $5\%$. 
It becomes larger in the region where the Hamiltonian 
with the $(-k)$-plaquette in the adjoint representation induces a phase
transition related to the $Z(2)$ symmetry. This region is indicated by the red shaded area in the figures. 
In principle only couplings in the weak coupling region above $\lambda =7$ are physically meaningful 
where the artificial $Z(2)$ symmetry in \eqnref{eqn:EffectiveLatHamiltonian} is spontaneously broken. 

The discussed analytical strong and weak coupling solutions \cite{pirner}
yield the following estimates for $\rho_0$ and $\delta_0$ in these limits
\bea
\rho_0&=&\left\{
\begin{array}{lcr}
 0 & \mathrm{for} & \lambda<<1 \\
 \sqrt{\lambda}~\gamma_{\tilde{\eta}}(\vec{0}) \phantom{~\tilde{\eta}^2} & \mathrm{for} & \lambda>>1
\end{array}
\right.\nonumber\\
\delta_0&=&\left\{
\begin{array}{lcr}
\frac{1}{3}~\lambda~\tilde{\eta}^2 & \mathrm{for} & \lambda<<1 \\
 \sqrt{\lambda}~\tilde{\eta}^2~\gamma_{\tilde{\eta}}(\vec{0}) & \mathrm{for} & \lambda>>1
\end{array}
\right.\nn\\
\gamma_{\tilde{\eta}}(\vec{0})&
\rightarrow& \frac{0.038}{\tilde{\eta}}~,~\mathrm{for}~\tilde{\eta} \rightarrow 0 ~.
\label{eqn:SupposedCoefficients}
\eea
The variationally determined parameters are in good agreement with the analytic predictions in the strong coupling regime which are represented by the dotted lines in \figref{fig:Params_first_16x16x16} and \figref{fig:Params_second_16x16x16}.
However, in the weak coupling regime the optimal parameters
differ from their analytical estimates \eqnref{eqn:SupposedCoefficients} which are indicated by the
arrows in the plots. Both analytic predictions disagree with the optimized values stronger
for decreasing values of $\tilde{\eta}$. This is natural, since the light cone limit
$\tilde{\eta} \rightarrow 0$ builds up correlations among plaquettes separated
along the longitudinal direction \cite{pirner}. The parameters optimizing
our product of single plaquette wave functionals can only effectively describe these
correlations.

In the physical relevant coupling region beyond $\lambda=7$,
it is possible to fit the variationally optimized wavefunctional parameters   
to the following parameterization
\bea
\rho_0(\lambda,\tilde{\eta}) &=& \sqrt{\lambda}~\gamma_1(\vec{0})~f_\rho(\lambda,\eta) \nn\\
\delta_0(\lambda,\tilde{\eta})& =& \sqrt{\lambda}~\gamma_1(\vec{0})~f_\delta(\lambda,\eta)\nn\\
f_i(\lambda,\tilde{\eta})& =& c_{0,i}\left[1+
\left(
\begin{array}{c}
c_{1,i} \\
c_{2,i} 
\end{array}
\right)
\cdot 
\left(
\begin{array}{c}
\lambda^{-1} \\
1-\tilde{\eta}
\end{array}
\right)\right. \nn\\
& & \left.~~~~~~~~~~~~~~~~~~~~~
+\onehalf
\left(
\begin{array}{c}
\lambda^{-1} \\
1-\tilde{\eta} 
\end{array}
\right)
\cdot 
\left(
\begin{array}{cc}
c_{3,i} & c_{4,i} \\
c_{4,i} & c_{5,i}
\end{array}
\right) 
\cdot
\left(
\begin{array}{c}
\lambda^{-1} \\
1-\tilde{\eta}
\end{array}
\right)\right]\nn\\
i&=& \rho,\delta~.
\label{eqn:fitquadraticform}
\eea
A good fit of the parameters
$c_{0,i},...,c_{5,i}$ minimizing $\chi^2$ in the range $\lambda\in[10,95]$ and
$\tilde{\eta}\in[0.15,1]$ is possible and yields the coefficients tabulated in 
\tabref{tab:fitcoeff}.
\begin{table}[h]
\center
\begin{tabular}{c|c|c|c|c|c|c}
i & $c_{0,i}$ & $c_{1,i}$ & $c_{2,i}$ & $c_{3,i}$ & $c_{4,i}$ & $c_{5,i}$ \\ \hline
$\rho$   & 0.90 & -1.74 &  0.72 &  8.12 & -0.40 & -0.27 \\ \hline
$\delta$ & 0.95 &  0.93 & -1.21 & -6.44 & -0.83 &  0.64 
\end{tabular}
\caption{Coefficients of eqn. (3.3) obtained from 
least square minimization. \label{tab:fitcoeff}}
\end{table}
The result of the fitting procedure \eqnref{eqn:fitquadraticform} is shown by the solid lines in 
\figref{fig:Params_first_16x16x16} and \figref{fig:Params_second_16x16x16}.
Having a parameterization of the ground state wavefunctional in dependence
of the nlc para\-meter $\tilde{\eta}$ at hand, we plan to extrapolate to the light cone
and finally calculate
hadronic cross sections by simulating how a color dipole moving near the light cone
hits a neutral hadron localized at $x^-=0$ \cite{Pirner:2002fe}. 
The color dipole can be represented by
a longitudinal-transversal Wilson loop elongated in $x^-$ direction and the simplified target by a transverse
plaquette. Varying the impact parameter one can sample the correlation function
of the two gauge-invariant objects and thereby obtain the profile function.
A necessary prerequisite of such a calculation for different velocities of the
dipole is that the lattice constant in transverse
direction stays constant for different $\tilde{\eta}$ values, in order to have a reliable
transverse length scale. 
\newline\newline 
{\bf Acknowledgements} 

We are grateful to the Max-Planck-Institut f\"ur Kernphysik Heidelberg for providing us
with resources on the Opteron cluster. D.~G. acknowledges funding by the European Union 
project EU RII3-CT-2004-506078 and the GSI Darmstadt. E.V.~P. thanks the Russian Foundation 
RFFI for the support in this work. E.-M.~I. is supported by DFG under contract FOR 465 (Forschergruppe
 Gitter-Hadronen-Ph\"anomenologie).


\begin{thebibliography}{99}

\bibitem{pauli}
  S.~J.~Brodsky, H.~C.~Pauli and S.~S.~Pinsky,
  Phys.\ Rept.\  {\bf 301} (1998) 299
  [arXiv:hep-ph/9705477].
  
\bibitem{Burkardt:2001jg}
M.~Burkardt and S.~Dalley,
Prog.\ Part.\ Nucl.\ Phys.\  {\bf 48} (2002) 317
[arXiv:hep-ph/0112007].
 
\bibitem{Bardeen:1979xx}
  W.~A.~Bardeen, R.~B.~Pearson and E.~Rabinovici,
  Phys.\ Rev.\  D {\bf 21} (1980) 1037.
  
\bibitem{Dalley:2003uj}
S.~Dalley and B.~van de Sande,
arXiv:hep-ph/0311368.


\bibitem{Pirner:1991im}
H.~J.~Pirner,
Prog.\ Part.\ Nucl.\ Phys.\  {\bf 29} (1992) 33.

\bibitem{Verlinde:1993te}
H.~Verlinde and E.~Verlinde,
arXiv:hep-th/9302104.
  
\bibitem{Arefeva:1993hi}
I.~Y.~Arefeva,
Phys.\ Lett.\ B {\bf 328} (1994) 411
[arXiv:hep-th/9306014].

\bibitem{Balitsky:2001gj}
  I.~Balitsky,
  arXiv:hep-ph/0101042.
  
\bibitem{Raufeisen:2004dg}
  J.~Raufeisen and S.~J.~Brodsky,
  Phys.\ Rev.\  D {\bf 70} (2004) 085017
  [arXiv:hep-th/0408108].

\bibitem{Prokhvatilov:1989eq}
  E.~V.~Prokhvatilov and V.~A.~Franke,
  Sov.\ J.\ Nucl.\ Phys.\  {\bf 49} (1989) 688
  [Yad.\ Fiz.\  {\bf 49} (1989) 1109].

\bibitem{pirner}
D. Gr\"unewald , E.-M. Ilgenfritz, E.V. Prokhvatilov and H.~J.~Pirner,
``Formulating Light Cone QCD on the Lattice,''
 work in progress.


\bibitem{Pirner:2002fe}
H.~J.~Pirner and F.~Yuan,
Phys.\ Rev.\ D {\bf 66} (2002) 034020
[arXiv:hep-ph/0203184].

\end{thebibliography}
\end{document}